\documentclass[aps,prb,twocolumn,showpacs,superscriptaddress,amsmath,amssymb,amsfonts,floatfix,longbibliography]{revtex4-2}

\usepackage[T1]{fontenc}
\usepackage{graphicx}
\usepackage{dcolumn}
\usepackage{listings}
\usepackage{xcolor}
\usepackage{multirow}
\usepackage[
  colorlinks,
  linkcolor=blue,
  filecolor=magenta,
  urlcolor=cyan,
  citecolor=blue
  ]{hyperref}%
\usepackage[normalem]{ulem}
\usepackage{siunitx}
\usepackage{braket}
\usepackage{extarrows}
\usepackage{bm}
\usepackage{nicematrix}

\newcommand{\ppx}[2][x]{\frac{\partial #2}{\partial #1}}

\begin{document}
\title{Meta-Altermagnetism and Enhanced Spin-Splitting } %
\author{Bogeng Wen}
    \affiliation{Department of Physics, University of Toronto, Toronto, Ontario M5S 1A7, Canada}
\author{Hae-Young Kee}
    \email{hy.kee@utoronto.ca}
    \affiliation{Department of Physics, University of Toronto, Toronto, Ontario M5S 1A7, Canada}
    \affiliation{Canadian Institute for Advanced Research, CIFAR Program in Quantum Materials, Toronto, Ontario M5G 1M1, Canada}
\date{\today}

\begin{abstract}
  Altermagnets are collinear magnets with zero net magnetization and spin-split electronic bands. 
  Here, we explore whether altermagnetism can exhibit a discontinuous jump, a phenomenon we refer to as meta-altermagnetism. We show that meta-altermagnetism can emerge in 2D systems when magnetic-order-induced band reconstruction drives spin-split van Hove singularities (vHSs) across the Fermi level. Using mean-field theory for an extended $t$-$J$ model on a two-sublattice square lattice including anisotropic hopping integral, we find that a first-order meta-altermagnetic transition produces an intermediate altermagnetic regime with reduced staggered magnetization but strongly enhanced magnetic octupole and time-reversal-odd spin-splitter conductivity. We further show that the first-order transition line terminates at a quantum critical endpoint, giving rise to a rich phase diagram as the chemical potential and anisotropic hopping strength are varied. Our results identify the spin split vHSs as a route to enhancing spin-splitter effect for spintronic applications.
\end{abstract}

\maketitle

\section{Introduction} \label{sec:intro}

Altermagnets are collinear antiferromagnets with zero net magnetization whose opposite-spin sublattices are related by a crystal rotation or mirror rather than by a translation or inversion~\cite{Y.Noda_2016_MomentumdependentBandSpin,S.Hayami_2019_MomentumDependentSpinSplitting,KH.Ahn_2019_AntiferromagnetismRuO2,LD.Yuan_2020_GiantMomentumdependentSpin,L.Smejkal_2020_CrystalTimereversalSymmetry,L.Smejkal_2022_ConventionalFerromagnetismAntiferromagnetism,L.Smejkal_2022_EmergingResearchLandscapea,T.Jungwirth_2025_AltermagnetismUnconventionalSpinordered,T.Jungwirth_2026_SymmetryMicroscopySpectroscopy}.
In the spin-group description~\cite{WF.Brinkman_1966_TheorySpinspaceGroups,L.Smejkal_2022_ConventionalFerromagnetismAntiferromagnetism,P.Liu_2022_SpinGroupSymmetryMagnetic,PA.McClarty_2024_LandauTheoryAltermagnetism,Z.Xiao_2024_SpinSpaceGroups,X.Chen_2024_EnumerationRepresentationTheory,Y.Jiang_2024_EnumerationSpinSpaceGroups}, this symmetry lifts the spin degeneracy of the bands in the absence of spin-orbit coupling and produces a momentum-dependent spin splitting of $d$-, $g$-, or $i$-wave form depending on the underlying structure.
The splitting has been reported in angle-resolved photoemission spectrum (ARPES) in MnTe~\cite{J.Krempasky_2024_AltermagneticLiftingKramers,S.Lee_2024_BrokenKramersDegeneracy}, CrSb~\cite{S.Reimers_2024_DirectObservationAltermagnetic,J.Ding_2024_LargeBandSplitting}, and in members of the layered vanadium oxychalcogenide family AV$_2$X$_2$O (A$=$K, Rb, Cs; X$=$Se, Te)~\cite{B.Jiang_2025_MetallicRoomtemperatureDwave,F.Zhang_2025_CrystalsymmetrypairedSpinvalleyLocking}, which belongs to a broader class of experimentally studied compounds built from 2D antiperovskite (inverse Lieb) layers~\cite{CC.Liu_2025_PhysicalPropertiesFirstprinciples,CC.Wei_2025_La2O3Mn2Se2CorrelatedInsulating,PH.Chang_2026_InverseLiebMaterials}.
Beyond these inorganic compounds, altermagnetism has also been proposed in the layered organic antiferromagnet $\kappa$-(BEDT-TTF)$_2$Cu[N(CN)$_2$]Cl ($\kappa$-Cl)~\cite{M.Naka_2019_SpinCurrentGeneration,S.Iguchi_2025_MagnetoopticalSpectraOrganic}.
First-principles calculations have further extended the candidates to 2D materials, such as monolayers of the van der Waals fluoride RuF$_4$\cite{J.Sodequist_2024_TwodimensionalAltermagnetsHigh,M.Milivojevic_2024_InterplayAltermagnetismWeak}.

Combining the vanishing net magnetization of antiferromagnets, which avoids stray fields and supports terahertz spin dynamics~\cite{K.Olejnik_2018_TerahertzElectricalWriting,C.Song_2025_AltermagnetsNewClass}, with the spin-split bands of ferromagnets, altermagnets have attracted growing attention in the field of spintronics~\cite{T.Jungwirth_2016_AntiferromagneticSpintronics,V.Baltz_2018_AntiferromagneticSpintronics,T.Jungwirth_2026_AltermagneticSpintronics} and are promising candidates for dense and fast magnetic random-access memory (MRAM)~\cite{B.Dieny_2020_OpportunitiesChallengesSpintronics,DF.Shao_2021_SpinneutralCurrentsSpintronics,C.Song_2025_AltermagnetsNewClass}.
Because of their spin polarized bands, altermagnetic (AM) metals can also convert a charge current into a spin-polarized current.
This charge-to-spin conversion, characterized by a time-reversal-odd spin conductivity, is known as the spin-splitter effect~\cite{M.Naka_2019_SpinCurrentGeneration, KH.Ahn_2019_AntiferromagnetismRuO2, R.GonzalezHernandez_2021_EfficientElectricalSpin, HY.Ma_2021_MultifunctionalAntiferromagneticMaterials, L.Smejkal_2022_GiantTunnelingMagnetoresistance}.
The associated spin-splitter torque has been reported in RuO$_2$~\cite{A.Bose_2022_TiltedSpinCurrent,H.Bai_2022_ObservationSpinSplitting,S.Karube_2022_ObservationSpinsplitterTorque}.
Such a torque has been used to switch a perpendicular magnetic tunnel junction without an external magnetic field, realizing a three-terminal MRAM device~\cite{Y.Guo_2025_MagneticMemoryDriven}.
Because the spin-splitter effect is carried by the spin-split Fermi surfaces, its efficiency depends on their geometry~\cite{JW.Lai_2025_DwaveFlatFermi}.

Since the spin-splitter effect depends on the magnitude of spin splitting, enhancing altermagnetism is important for spintronic applications. In metallic altermagnets, spin splitting arises from the interplay between antiferromagnetic (AFM) order and anisotropic hopping\cite{S.Hayami_2019_MomentumDependentSpinSplitting,PA.McClarty_2024_LandauTheoryAltermagnetism,M.Roig_2024_MinimalModelsAltermagnetism}. A natural question is whether the strength of antiferromagnetic order determines the magnitude of AM spin splitting. Here, we explore whether altermagnetism can be enhanced by tuning the chemical potential or anisotropic hopping integral, focusing on the possibility of a discontinuous jump in spin splitting, which we refer to as \emph{meta-altermagnetism}, in analogy with metanematic transitions.

Nematic order, associated with spontaneous rotational symmetry breaking through a Pomeranchuk instability\cite{IY.Pomeranchuk_1959_StabilityFermiLiquid,H.Yamase_2000_InstabilityFormationQuasiOneDimensional,CJ.Halboth_2000_WaveSuperconductivityPomeranchuk,V.Oganesyan_2001_QuantumTheoryNematic,E.Fradkin_2010_NematicFermiFluids}, can undergo a first-order transition near van Hove singularities (vHSs)\cite{I.Khavkine_2004_FormationElectronicNematic,H.Yamase_2005_MeanfieldTheorySymmetrybreaking}. Moreover, discontinuous transitions can occur within the nematic phase without additional symmetry breaking, a phenomenon known as a metanematic transition\cite{C.Puetter_2007_MetanematicTransitionsBilayer,S.Raghu_2009_MicroscopicTheoryNematic}. A spin-triplet Pomeranchuk instability can also generate opposite nematic distortions in the two spin sectors\cite{V.Oganesyan_2001_QuantumTheoryNematic,C.Wu_2007_FermiLiquidInstabilities}, giving rise to a nematic-spin-nematic state that shares the characteristic symmetry of altermagnetism without requiring primary AFM order\cite{C.Wu_2007_FermiLiquidInstabilities,CM.Puetter_2010_MicroscopicRouteNematicity,SK.Das_2025_LocalSpinNematicity,T.Jungwirth_2025_AltermagnetismUnconventionalSpinordered,Z.Liu_2026_AltermagnetismSuperconductivityShort}. Motivated by these connections, we investigate whether meta-altermagnetism can emerge in 2D systems, leading to a discontinuous enhancement of spin splitting. To this end, we consider an extended $t$-$J$ model on a square lattice with two sublattices.

Using a mean field theory, we find three AM regimes, denoted by AM1, AM2, and AM1$'$. They are separated by first-order transition accompanying a discontinuous jump in the altermagnetism suggesting the meta-altermagnetic transitions.
We will show that they are associated with the spin-split vHS crossing the Fermi level.
For example, across the AM1-AM2 transition, the staggered moment $m$ decreases while the induced $d$-wave magnetization $m_d$ is strongly enhanced.
A separate meta-altermagnetic line terminates at a quantum critical endpoint (QCEP), beyond which the reconstruction becomes a continuous crossover.
We employed a local Landau-Ginzburg expansion of the free energy about a finite moment to describe these transitions and their endpoints.
The time-reversal-odd spin conductivity was found to be strongly enhanced in AM2 region and changes discontinuously across the first-order meta-altermagnetic transition, providing a transport signature of the reconstruction.

The rest of this paper is organized as follows.
In Sec.~\ref{sec:model}, we introduce the two-sublattice $t$-$J$ model and its mean-field decoupling, which generates AFM order $m$ and permits $d$-wave magnetism $m_d$ through anisotropic hopping.
In Sec.~\ref{sec:pd}, we %
present the zero-temperature phase diagram, and relate the boundaries between the AM regimes AM1, AM2, and AM1$'$ to the Fermi level crossing the spin-split vHS.
In Sec.~\ref{sec:landau}, we expand the free energy about a finite staggered moment and use this local Landau theory to determine where a first-order meta-altermagnetic line terminates at a critical endpoint.
In Sec.~\ref{sec:spin-splitter}, we compute the time-reversal-odd spin conductivity and show that it follows $m_d$, being strongly enhanced in AM2.
Section~\ref{sec:summary} summarizes our results.

\section{Mean-Field Model} \label{sec:model}

\begin{figure}[t]
  \centering
  \includegraphics[width=\columnwidth]{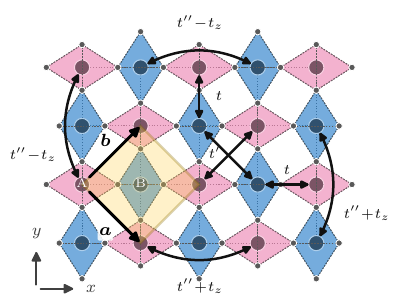}
  \caption{
    Lattice structure of our model, shown with rutile MnF$_2$ viewed down the $c$ axis, with each MnF$_6$ octahedron drawn in projection and the F atoms as grey dots.
    The Mn sites at heights $0$ and $c/2$ form two sublattices $A$ and $B$ (pink and blue), exchanged by the $4_2$ screw, which rotates their octahedral environments by $90^\circ$.
    The shaded cell, spanned by $\bm{a}$ and $\bm{b}$, contains one site of each sublattice.
    Arrows mark the hoppings of Eq.~\eqref{eq:coeffs}: the nearest-neighbour hopping $t$, the second-neighbour hopping $t'$ along the cell diagonals, and the third-neighbour hoppings $t''\pm t_z$, whose strong and weak directions interchange between the sublattices.
    We idealize the weakly rectangular layer to a square lattice ($|\bm{a}|=|\bm{b}|$) to simplify the analysis.
  }
  \label{fig:model}
\end{figure}

We consider a 2D $t$-$J$ model on a square lattice with two sublattices $A$ and $B$ (see Fig.~\ref{fig:model}), following\cite{TA.Maier_2023_WeakcouplingTheoryNeutron},
\begin{equation}
  H = H_{0} + H_{J} = \sum_{\mathbf{k}}\Psi_{\mathbf{k}}^{\dagger}\, h(\mathbf{k})\,\Psi_{\mathbf{k}} + J\sum_{\langle ij\rangle}\vec{S}_i\cdot\vec{S}_j,
  \label{eq:H}
\end{equation}
where the spinor $\Psi_{\mathbf{k}}=(c_{\uparrow A},c_{\uparrow B}, c_{\downarrow A},c_{\downarrow B})^{T}$ is ordered spin first ($\sigma\otimes\tau$), with Pauli matrices $\sigma_i$ and $\tau_i$ acting on the spin and sublattice spaces, respectively.
The single-particle Bloch Hamiltonian reads
\begin{equation}
  h(\bf{k})=\epsilon_0(\bf{k})\mathbb{I} + f_z(\bf{k})\tau_z + f_x(\bf{k})\tau_x,
  \label{eq:hk0}
\end{equation}
where
\begin{equation}
    \begin{aligned}
          \epsilon_0 ({\bf k}) &= -4t'\cos\tfrac{k_x}{2}\cos\tfrac{k_y}{2} - 2t''(\cos k_x+\cos k_y), \\
          f_z({\bf k})        &= -2t_z(\cos k_x-\cos k_y), \\
          f_x({\bf k})      &= -2t (\cos\tfrac{k_x}{2}+\cos\tfrac{k_y}{2}).
    \end{aligned}
    \label{eq:coeffs}
\end{equation}
Here $t$ denotes the nearest-neighbour inter-sublattice ($A$--$B$) hopping, $t'$ the second-neighbour intra-sublattice hopping along the cell diagonals (i.e., $\bm{a}$ and $\bm{b}$), $t''$ the third-neighbour intra-sublattice hopping along $2\hat{\bm{x}}$ and $2\hat{\bm{y}}$, and  $t_z$ anisotropic $d$-wave ($B_{1g}$) hopping that alternates between the two sublattices, as shown in Fig.~\ref{fig:model}.

This minimal tight-binding model represents
a single layer of rutile MnF$_2$ (space group $P4_2/mnm$)~\cite{M.Roig_2024_MinimalModelsAltermagnetism}, with a Bloch Hamiltonian possessing tetragonal point-group symmetry $D_{4h}$. For $t_z=0$, the two sublattices have identical hopping environments, and an effective translation interchanges them. Combined with a $\pi$ spin rotation $C_2$, this translation enforces spin degeneracy at each wavevector, yielding a conventional antiferromagnet. For $t_z\neq0$, the alternating hoppings $t''\pm t_z$ break this effective translation, while the fourfold rotation $C_{4z}$ still interchanges the sublattices. The spin group therefore contains $[C_2\parallel C_{4z}]$, allowing AM spin splitting. The corresponding term $f_z=-2t_z(\cos k_x-\cos k_y)$ transforms as $B_{1g}$, giving rise to $d$-wave altermagnetism with a $d_{x^2-y^2}$ form factor. Since both $f_z$ and $\tau_z$ transform as $B_{1g}$, their product $f_z\tau_z$ belongs to the trivial representation $A_{1g}$, allowing $t_z$ to couple to in-plane biaxial strain $\varepsilon_{xx}+\varepsilon_{yy}$.

The same Hamiltonian can describe lower-symmetry systems such as monolayer RuF$_4$ (space group $P2_1/c$)~\cite{J.Sodequist_2024_TwodimensionalAltermagnetsHigh,M.Milivojevic_2024_InterplayAltermagnetismWeak,S.Zeng_2024_DescriptionTwodimensionalAltermagnetism}. In this case, $|\bm a|\neq|\bm b|$ introduces an additional anisotropic second-neighbor hopping term proportional to the identity matrix, which does not directly contribute to spin splitting. Alternatively, the anisotropy $t_z$ can arise from orbital ordering rather than lattice distortion. An example is staggered $d_{xz}/d_{yz}$ orbital order coexisting with Néel order~\cite{V.Leeb_2024_SpontaneousFormationAltermagnetism}, as proposed for Sr$_2$CrO$_4$~\cite{QN.Meier_2026_NetCompensatedAltermagnetism}.

For the interaction part, we perform a standard Hartree-Fork decoupling:
\begin{equation}
\begin{aligned}
     J\vec{S}_A\cdot\vec{S}_B  \xrightarrow{\rm MF}  &\frac{J}{2}\big[\langle\vec{S}_A+\vec{S}_B\rangle\cdot(\vec{S}_A+\vec{S}_B) \\
      &-\langle\vec{S}_A-\vec{S}_B\rangle\cdot(\vec{S}_A-\vec{S}_B)\big] - J\langle\vec{S}_A \rangle \cdot \langle\vec{S}_B\rangle,
\end{aligned}
\end{equation}
which yields an effective attractive interaction in the AFM channel, characterized by the order parameter $\langle \vec S_A\rangle=-\langle\vec S_B\rangle \equiv m {\hat n}$, within mean-field theory: 
\begin{equation}
  V_{\rm MF}=-2Jm\sum_{\mathbf{k}}\Psi_{\mathbf{k}}^{\dagger}\sigma_z\tau_z\Psi_{\mathbf{k}} + 4N_k Jm^2 .
\end{equation}

Combining $V_{\rm MF}$ and the tight-binding Hamiltonian $H_0$ we get the mean-field Hamiltonian $H_{\rm MF} = H_0 + V_{\rm MF}$, whose eigenvalues are given by:
\begin{equation}
  \varepsilon_{\pm,\sigma}(\mathbf{k};m)=\epsilon_0(\mathbf{k}) \pm \sqrt{f_x({\bf k})^2+(f_z ({\bf k}) -2Jm\sigma)^2},
  \label{eq:eigenvalues}
\end{equation}
where $\sigma = \pm 1$ labels the spin.

The staggered moment $m$ 
together with a finite $t_z$ 
generates the spin-splitting
characterized by the AM  $d$-wave magnetization $m_d$:
\begin{equation}
  m_d=\frac{1}{N_k}\sum_{n, \mathbf{k}} \left(\cos k_x-\cos k_y\right) \braket{S_z}_{n, \mathbf{k}}  n_F(\varepsilon_{n, \mathbf{k}} - \mu),
  \label{eq:md}
\end{equation}
where $n_F$ is the Fermi distribution, $(\cos k_x-\cos k_y)$ is the $B_{1g}$ ($d$-wave) form factor.
A finite $m_d$ measures the imbalance between the spin-up and spin-down Fermi surfaces.
Below we present the phase diagram showing the emergence of $m$ and $m_d$  as functions of $t_z$ and $\mu$.

\section{Phase Diagram and Meta-Altermagnetic Transitions} \label{sec:pd}

\begin{figure}[!h]
  \centering
  \includegraphics[width=\columnwidth]{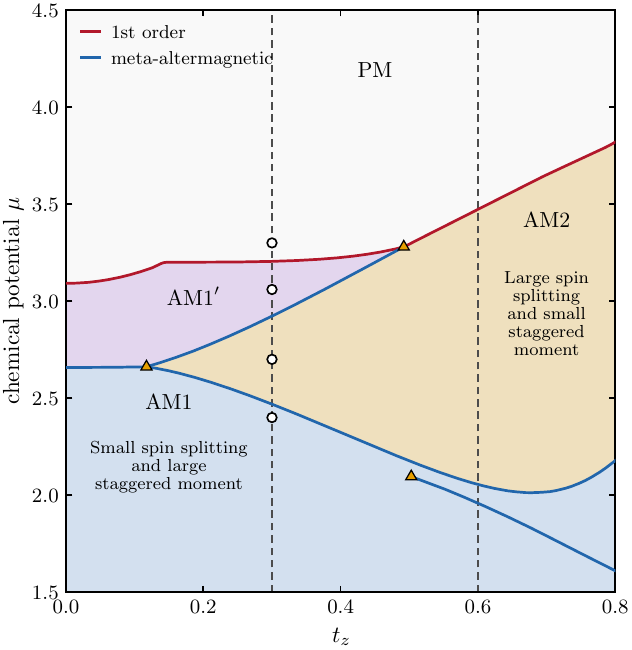}
  \caption{
    Zero-temperature phase diagram over $0\leq t_z\leq0.8$ and $1.5\leq\mu\leq4.5$ for $t=-1$, $t'=0.4$, $t''=0.8$, and $J=4.5$.
    Shading denotes PM and the three AM regimes AM1, AM2, and AM1$'$.
    Red curves are PM-AM boundaries, all first order in this range; blue curves are internal meta-AM boundaries.
    Triangles mark the two triple points and the QCEP.
    In AM1 (AM1$'$), both spin-split vHS lie above (below) $E_F$, while in AM2 $E_F$ has crossed only the lower one (see Fig.~\ref{fig:bands}).
    Both AM1 and AM1$'$ connect continuously to the $t_z=0$ antiferromagnet (AFM). 
    The dashed vertical lines are the $t_z=0.3$ and $t_z=0.6$ cuts scanned in Fig.~\ref{fig:mu-scan}, and the four open circles on the $t_z=0.3$ cut are the representative points whose band structures and Fermi surfaces are shown in Fig.~\ref{fig:bands}.
  }
  \label{fig:phase-diagram}
\end{figure}

By minimizing the mean-field free energy with respect to $m$ at zero temperature and fixed $J=  4.5|t|$, we obtain the phase diagram shown in Fig.~\ref{fig:phase-diagram}, as a function of the anisotropic hopping $t_z$ and chemical potential $\mu$.

At $t_z=0$, as $\mu$ is lowered from the paramagnetic (PM) phase, a spin-degenerate antiferromagnetic state with a small staggered moment first emerges, continuously connected to the AM1$^\prime$ phase (purple region) at finite $t_z$. Upon further decreasing $\mu$, the staggered moment jumps discontinuously to a larger value across a first-order transition, entering a state continuously connected to the AM1 phase (blue region) at finite $t_z$.

Note that at $t_z=0$, the AFM state is spin-degenerate due to an effective symmetry. Turning on $t_z$ breaks this effective symmetry and allows AM spin splitting. However, this does not introduce additional spontaneous symmetry breaking, since $t_z$ is a parameter of the Hamiltonian.
Thus, within each of the purple and blue regions, the AFM state at $t_z=0$ is adiabatically connected to the corresponding AM state at finite $t_z$, without a phase transition. 

The AM1-AM1$'$ line stays nearly flat in $\mu$ up to a triple point at $t_z\simeq0.12$, where it splits into two meta-altermagnetic lines that enclose an intermediate regime, which we call AM2, where the large spin-splitting emerges. We show the behaviors of $m$, $m_d$, and total density $n$ as a function of $\mu$ for fixed values of $t_z =0.3$ and $0.6$ indicated by the dashed line in Fig. \ref{fig:phase-diagram}

\begin{figure}[!h]
  \centering
  \includegraphics[width=\columnwidth]{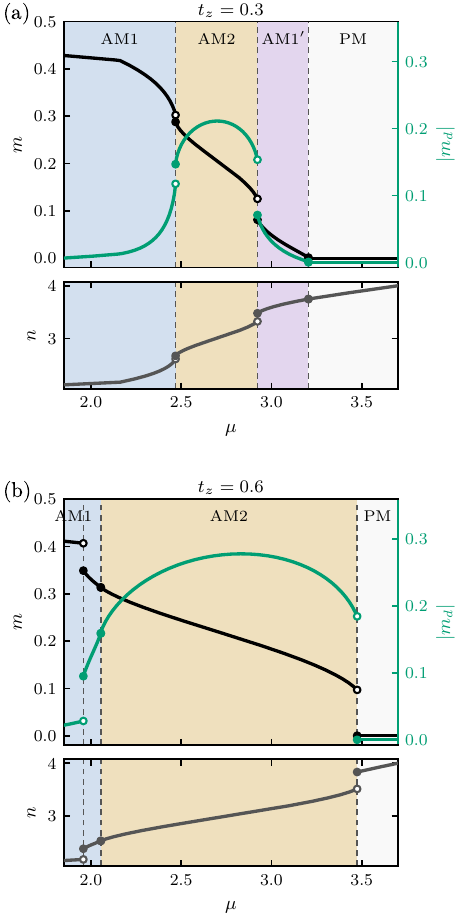}
  \caption{
    The staggered moment $m$ (black), $d$-wave magnetization $|m_d|$ (green), and filling $n$ as functions of $\mu$ at $T=0$ for $t=-1$, $t'=0.4$, $t''=0.8$, and $J=4.5$.
    (a) At $t_z=0.3$, the cut crosses AM1, AM2, AM1$'$, and PM, with first-order transitions;
    (b) At $t_z=0.60$, the cut crosses AM1, AM2, PM, with first-order transitions. A meta-altermagnetic transition happens within the AM1 region.
  }
  \label{fig:mu-scan}
\end{figure}

\begin{figure*}[t]
  \centering
  \includegraphics[width=\textwidth]{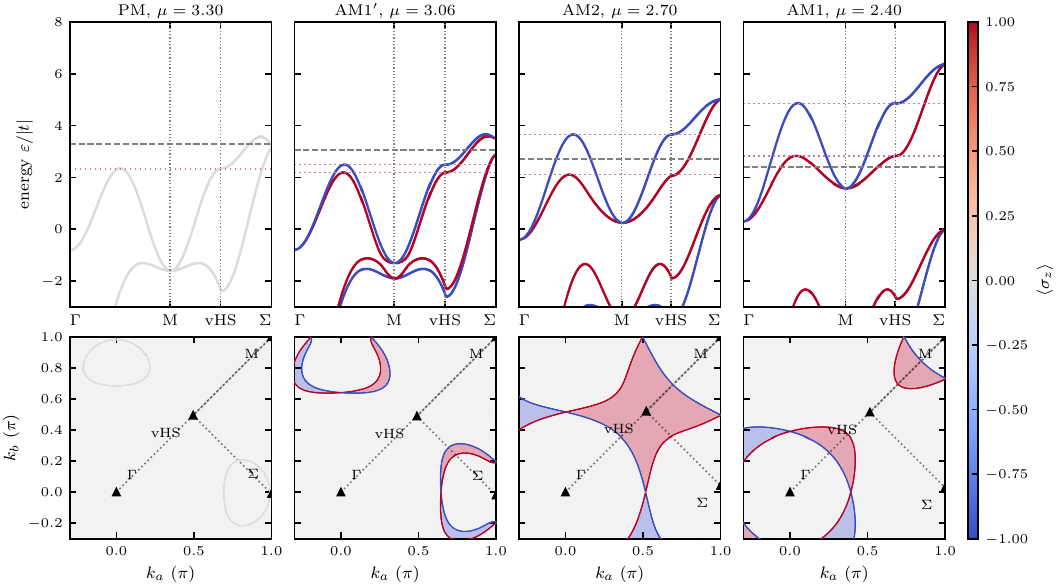}
  \caption{
    Band structure $\varepsilon_{\pm,\sigma}(\bm k)$ (top) and Fermi surface (bottom) along $\Gamma$-$M$-$\rm vHS$-$\Sigma$ at representative points of the four regimes; the path is drawn on each Fermi-surface panel.
    Momenta are given along the cell vectors $\bm a$ and $\bm b$ of Fig.~\ref{fig:model}, $k_a=(k_x-k_y)/2$ and $k_b=(k_x+k_y)/2$.
    $\rm vHS$ is a spin-split, symmetry-unprotected saddle point on a $\Gamma$-$M$ line, so its momentum moves from panel to panel; $\Sigma$ is its projection onto the zone boundary.
    The four values of $\mu$ are the open circles marked on the $t_z=0.3$ cut of Fig.~\ref{fig:phase-diagram}, ordered from left to right as they run from top to bottom along that cut.
    Bands are coloured by spin; the dashed line is the Fermi level and the dotted lines the two van Hove singularities.
    In AM1(AM1$'$), both vHS lie above(below) $E_F$; in AM2 $E_F$ sits between the two vHS, enlarging the splitting between the two Fermi surfaces of different spin; in PM, the two vHS are degenerate.
  }
  \label{fig:bands}
\end{figure*}

Fig.~\ref{fig:mu-scan} shows two chemical-potential cuts through the phase diagram.
At $t_z=0.3$ (Fig.~\ref{fig:mu-scan}(a)), both $m$, $|m_d|$, and charge density $n$ jump at the AM1-AM2 boundary $\mu\simeq2.47$ and again at the AM2-AM1$'$ boundary $\mu\simeq2.92$, before the first-order transition to PM at $\mu\simeq3.20$.
When entering the AM2 regime from AM1, the staggered moment $m$ decreases while $|m_d|$ is strongly enhanced, demonstrating that a larger staggered moment does not imply a larger momentum-dependent spin splitting.
This is in agreement with the Fermi-surface reconstruction shown in Fig.~\ref{fig:bands}.
At $t_z=0.6$ (Fig.~\ref{fig:mu-scan}(b)), $m$, $|m_d|$, and $n$ jump at the meta-altermagnetic line inside AM1 at $\mu\simeq1.96$, shortly before the cut enters AM2 at $\mu\simeq2.06$, and the trend that $m$ decreases and $|m_d|$ grows still persists, as the consequence of Fermi-surface reconstruction caused by the vHS.
This cut does not cross AM1$'$ and leaves AM2 through a first-order transition to PM at $\mu\simeq3.47$; the suppressed $m$ and $|m_d|$ in AM1$'$ is demonstrated in the $t_z=0.3$ cut.
A kink occurs on $m$, $|m_d|$, and $n$ at $\mu\simeq2.16$ inside AM1 of the $t_z=0.3$ cut, which corresponds to a Lifshitz transition at the band bottom at $M$-point in the Brillouin zone (BZ) beyond the QCEP, where the pocket at $M$ opens without a discontinuity, as we will show below.
In the $t_z=0.6$ cut, by contrast, $M$-point turns into a saddle point, and the same Lifshitz transition becomes the first-order jump at $\mu\simeq1.96$.

Three local minima in the mean-field free energy coexist around the triple point.
The lower AM1-AM2 line extends over the whole $t_z$ range shown, whereas the upper AM2-AM1$'$ line reaches the PM boundary at a second triple point, beyond which AM2 borders PM directly.

The three AM regimes are associated with the vHSs in the band structure. They are separated by the Fermi level crossing the spin-split vHS on the $\Gamma$-$M$ lines, which originate from the PM vHS present for $t_z<\sqrt{t''^2-t'^2/4}$. Related band structures are shown in Fig.~\ref{fig:bands}.
In AM1, both vHS lie above the Fermi level, whereas in AM1$'$ both lie below it; the Fermi surfaces of the two spins therefore differ less in both regimes, and in AM1$'$ $m_d$ is suppressed together with the staggered moment $m$.
In the AM2 region, the chemical potential has crossed only the lower spin-split vHS, which strongly enhances the deformation of the Fermi surface, and hence the $d$-wave magnetization $m_d$, compared to the AM1 region.
AM2 also hosts the most favorable conditions for ordering.
Considering the critical coupling $J_c$, the smallest $J$ where the AM ordering sets in, we find that the AM2 region starts to order at $J_c\approx0.21$ on one particular parameter point, where a nearly flat maximum of the upper band sits close to the Fermi level.
The surrounding AM1 only begins to order at $J\approx2.75$, at which about $30\%$ of AM2 is already ordered.

The meta-altermagnetic transitions between these regimes are associated with a spin-split vHS crossing the Fermi level, through the same mechanism as the first-order meta-nematic transition in bilayer systems near van Hove filling~\cite{C.Puetter_2007_MetanematicTransitionsBilayer}. %
Such a metanematic transition can be understood in terms of successive nematic transitions in multiband systems. In a 2D bilayer system, it has been shown that once nematic order develops, both bands experience broken rotational symmetry. However, as the chemical potential varies and one of the Fermi surfaces reaches the BZ boundary, the nematic order parameter undergoes additional discontinuous jump, resulting in a metanematic transition.

It is also important to recall that the nematic transition can become first-order, preempting the continuous Pomeranchuk instability expected from a conventional Fermi-liquid analysis that neglects the effects of vHSs on a lattice\cite{I.Khavkine_2004_FormationElectronicNematic}. To examine this preemption, we investigate the free energy as a function of the chemical potential $\mu$ and identify where its second derivative diverges.

\begin{figure}[!tp]
  \centering
  \includegraphics[width=\columnwidth]{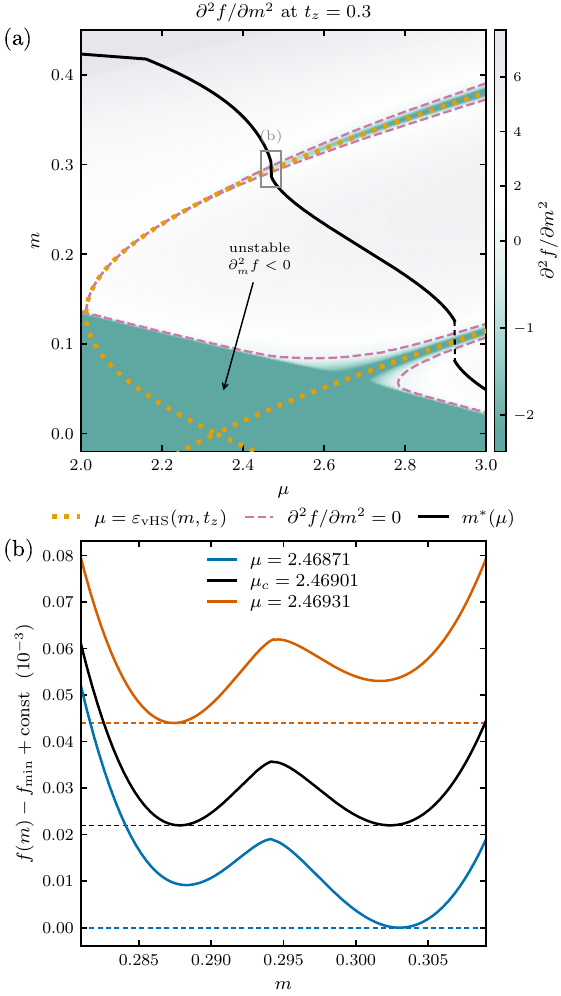}
  \caption{
    Free-energy landscapes at $T=0$: (a),(b) $t_z=0.3$, with a first-order meta-altermagnetic transition.
    The normalization is per magnetic site as in Eq.~\eqref{eq:landau-full-free-energy}.
    (a) $\partial_m^2 f$ with the equilibrium branch $m^\ast(\mu)=\arg\min_m f(m,\mu)$ (solid black), broken at the first-order transitions and bridged there by dashed segments.
    The purple dashed lines are $\partial_m^2 f=0$ contours, and the orange dotted lines trace the spin-split vHS $\mu=\varepsilon_{\rm vHS}(m,t_z)$.
    The grey rectangle indicates the region sampled by the free-energy cuts.
    (b) $f(m)$ near the coexistence value $\mu_c\simeq2.47$, showing two minima exchanging stability.
    Each curve is measured from its own minimum marked by the dashed lines, and offset vertically for comparison.}
  \label{fig:free-energy}
\end{figure}

In our case, the two spin sectors exhibit distinct band structures. As the chemical potential varies, the staggered magnetization $m$ changes, and a discontinuous jump occurs in one of the spin sectors, giving rise to meta-altermagnetism. To examine this transition, we investigate the free energy as a function of the AFM order parameter $m$ and identify where its second derivative vanishes, signaling a continuous instability that may be preempted by a first-order transition. 
The free energy per site is given by
\begin{equation}
 f(m)=2Jm^2-\frac{T}{4}\sum_{n,\sigma}\int_{\rm BZ}\frac{\mathrm{d}^2k}{(2\pi)^2}\ln\left[1+\mathrm{e}^{\beta(\mu-\varepsilon_{n,\sigma}(\mathbf{k};m))}\right],
 \label{eq:landau-full-free-energy}
\end{equation}
where $n=\pm$ labels the two bands, and $\sigma=\pm1$ labels spin, BZ denotes the magnetic BZ, and $\beta=1/T$. 
Its curvature $\partial_m^2 f$ is shown in Fig.~\ref{fig:free-energy}(a).
The $\mu = \varepsilon_{\rm vHS}$ trajectories are marked by orange dotted lines, and the equilibrium staggered moment $m^\ast(\mu) = \arg\min_m f(m,\mu)$ is shown as a black line.
Narrow regions of negative curvature, bounded by the $\partial_m^2 f=0$ contours, follow these trajectories, and $m^\ast$ jumps across them at the meta-altermagnetic transitions.
As shown in Fig.~\ref{fig:free-energy}(a), the first-order transition occurs at $\mu_c$, where two distinct values of $m$ have the same free energy, as illustrated in Fig.~\ref{fig:free-energy} (b).
Interestingly, while the staggered magnetization $m$ decreases, the $d$-wave magnetic order parameter $m_d$ exhibits a discontinuous increase across AM1 to AM2.

A separate meta-altermagnetic line lies inside AM1 and terminates at a quantum critical endpoint (QCEP), beyond which the two sides of the line are connected continuously, reflecting its non-symmetry-breaking nature as shown in Fig.~\ref{fig:phase-diagram}.
At large $t_z$, a spin-degenerate vHS appears at the zone corner $M$ on the upper bands, where additional spin-group symmetry $[C_2\parallel M_{a/b}]$ forbids spin splitting.
This vHS is associated with the meta-altermagnetic transition in the same way as those between AM1, AM2, and AM1$'$.
With decreasing $t_z$, $M$ turns into a nearly flat band bottom, whose large density of states keeps the line first order down to the QCEP as the band bottom crosses the Fermi level.
This last stage is no longer controlled by a logarithmic singularity and can be described by a Landau expansion of the free energy about a finite moment, which we develop in the next section.

\section{Landau theory and critical end point on meta-altermagnetic transition} \label{sec:landau}

To describe the meta-altermagnetic transitions and their critical endpoints, we develop a local Landau expansion of the grand potential per magnetic site $f(m)$.
As the nature of meta-altermagnetic transition, there is no symmetry broken occurring in this process, but instead there is only the magnetic amplitude changed.
We therefore expand on a finite reference moment $m_0>0$, where $m_0$ is already one of the saddle points of the free energy function. 
Odd powers of $\delta m=m-m_0$ are now allowed despite the global symmetry $f(m)=f(-m)$, but the linear term vanishes:
\begin{equation}
 f(m_0+\delta m)= f_0+\frac{a}{2}(\delta m)^2+\frac{b}{3}(\delta m)^3+\frac{c}{4}(\delta m)^4 + \cdots ,
 \label{eq:landau-local}
\end{equation}
with $f_0=f(m_0)$ and
\begin{equation}
 a=\left.\partial_m^2 f\right|_{m_0},\quad
 b=\frac12\left.\partial_m^3 f\right|_{m_0},\quad
 c=\frac16\left.\partial_m^4 f\right|_{m_0}.
 \label{eq:landau-coefficients}
\end{equation}
All three coefficients depend on $\mu$, $t_z$, $T$, and the expansion center $m_0$.
Note that this is a local description with near an ordinary endpoint, not a global quartic approximation to the even function $f(m)$.
Hence the free energy polynomial is kept under fourth order, whereas a global even polynomial requires at least eighth order to accommodate two distinct positive-moment interior minima and the intervening maximum.

For $c>0$, the stationary condition $\partial_m f=\delta m\,[a+b\,\delta m+c(\delta m)^2]=0$ admits two local minima separated by a maximum when satisfy $a\neq0$ and the discriminant: 
\begin{equation}
 \Delta=b^2-4ac > 0.
 \label{eq:landau-discriminant}
\end{equation}
Within the range where the local polynomial description Eq.\eqref{eq:landau-local} captures the competing free energy $f(m)$, the discriminant therefore divides parameter space into a two-minimum region with $\Delta>0$ and a single-minimum region with $\Delta<0$.
The former allows a first-order transition when the two minima at $0<m_1<m_2$ become equally deep global minima, $f(m_1)=f(m_2)=\min_m f(m)$, whereas the latter cannot support a jump between these states.
The loss of the double-well structure thus allows the first-order line to terminate at a critical endpoint.

In addition to \eqref{eq:landau-discriminant}, another discriminant determines when $f(m_2) <  f(m_1)$ could happen by:
\begin{equation}
    \Delta' =  b^2 - \frac{9}{2} ac > 0 .
    \label{eq:transition-discriminant}
\end{equation}
Within the region that \eqref{eq:landau-discriminant} is satisfied, $\Delta'=0$ sets when the meta-altermagnetic transition could happen.
When both $\Delta=0$ and $\Delta'=0$ happen, the meta-altermagnetic transition terminates, and thus generate a quantum critical end point (QCEP) at $T=0$, which gives $a=b=0$ with $c>0$. Both minima and the intervening maximum merge at QCEP. The transition becomes continuous after QCEP.

\section{Spin-Splitter Conductivity}
\label{sec:spin-splitter}

Electrical generation of spin-polarized currents provides a route to manipulate magnetic order in spin-transfer and spin-orbit-torque devices.
Altermagnets offer a nonrelativistic mechanism for this charge-to-spin conversion because their spin-up and spin-down bands can have different shapes.
At the mean-field level, the conductivity of each spin channel follows from the standard semiclassical expression:
\begin{equation}
  \begin{aligned}
    \sigma^s_{ij}
    &=-e^2\tau\sum_n\int_{\rm BZ}\frac{d^2k}{(2\pi)^2}
      v_{i,n,s}(\bm k)v_{j,n,s}(\bm k) \left.\ppx[\varepsilon]{n_F}\right|_{\varepsilon=\varepsilon_{n,s}(\bm k)} \\
    &=-\frac{e^2\tau}{\hbar^2}\sum_n\int_{\rm BZ}\frac{d^2k}{(2\pi)^2}
      \ppx[k_i]{\varepsilon_{n,s}}\ppx[k_j]{\varepsilon_{n,s}} 
      \left.\ppx[\varepsilon]{n_F}\right|_{\varepsilon=\varepsilon_{n,s}(\bm k)} .
  \end{aligned}
  \label{eq:spin-resolved-conductivity}
\end{equation}
Here $i,j$ are the current and electric-field directions, $n$ labels the bands and $s=\uparrow,\downarrow$ labels the spins.
$\tau$ represents the transport lifetime, for which we chose a unitless $\tau = 1$ for numerical calculation.
$v_{i,n,s} = \hbar^{-1} \partial_{k_i} \varepsilon_{n,s}$ is the group velocity, BZ denotes the Brillouin zone.
$n_F$ is te Fermi-Dirac function $n_F(\varepsilon) = (\exp[(\varepsilon-\mu)/T]+1)^{-1}$ with $k_B=1$.
Eq.~\eqref{eq:spin-resolved-conductivity} shows directly that differently shaped spin-up and spin-down bands generally give $\sigma^\uparrow_{ij}\neq\sigma^\downarrow_{ij}$; the resulting current is spin polarized, which is described as the spin-splitter effect \cite{R.GonzalezHernandez_2021_EfficientElectricalSpin}.

In the square-lattice $(x,y)$ frame, the AM spin splitting transforms according to the $B_{1g}$ irreducible representation and behave as $d_{x^2-y^2}$, or $d_{ab}$ in the $45^\circ$-rotated magnetic-cell axes.
The space group operation together with $C_2$ spin rotation exchanges $x\leftrightarrow y$ together with $\uparrow\leftrightarrow\downarrow$, and therefore
\begin{equation}
  \sigma^\uparrow_{xx}=\sigma^\downarrow_{yy} \neq \sigma^\uparrow_{yy}=\sigma^\downarrow_{xx}.
  \label{eq:spin-conductivity-c4}
\end{equation}
Moreover, the mirror symmetry $M_x$ or $M_y$ ensures that $\sigma^s_{yx}=\sigma^s_{xy}=0$ for both spin channels.
For the positive AFM vector domain used below, the calculated ordering is $\sigma^\uparrow_{xx}=\sigma^\downarrow_{yy}>\sigma^\downarrow_{xx}=\sigma^\uparrow_{yy}$.
Reversing the AFM vector reverses the inequality.
Thus an electric field along either $x$ or $y$ produces a longitudinal spin-polarized current, with opposite spin polarization in the two directions.

When rotated into the magnetic-cell unit vectors (see Fig.~\ref{fig:model}), all longitudinal components become equal, whereas the transverse components have opposite signs for the two spin channels: $\sigma^\uparrow_{ab}=-\sigma^\downarrow_{ab}$.
Consequently, an electric field along $b$ generates a pure spin current along $a$ with no accompanying transverse charge current, which can be characterized by the time-reversal-odd spin conductivity\cite{R.GonzalezHernandez_2021_EfficientElectricalSpin}, expressed in charge-current units:
\begin{equation}
  \sigma^{\mathrm{odd},z}_{ab} \equiv\sigma^\uparrow_{ab}-\sigma^\downarrow_{ab} =2\sigma^\uparrow_{ab}=\sigma^{\mathrm{odd},z}_{ba},
  \label{eq:odd-spin-conductivity}
\end{equation}

\begin{figure*}[t]
  \centering
  \includegraphics[width=\textwidth]{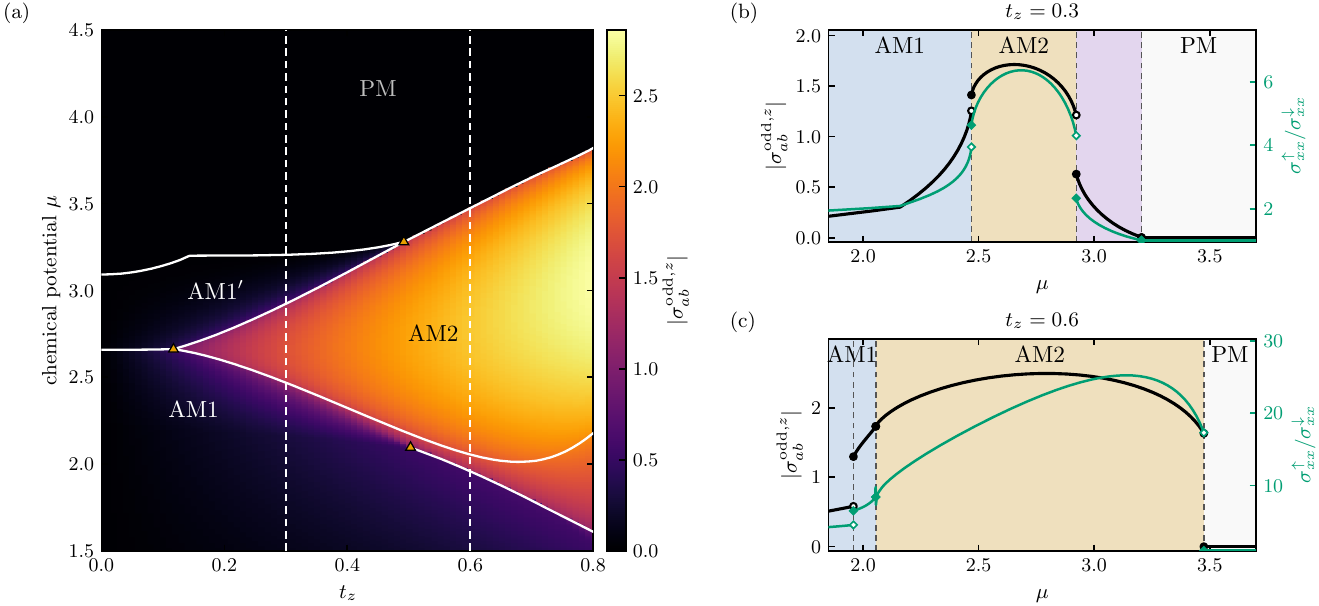}
  \caption{
    Spin-splitter effect computed at $T=0$ for $t=-1$, $t'=0.4$, $t''=0.8$, and $J=4.5$.
    (a) Map of $|\sigma^{\mathrm{odd},z}_{ab}|$. The boundary conventions follow Fig.~\ref{fig:phase-diagram}.
    (b), (c) Scan of $\sigma^{\mathrm{odd},z}_{ab}$ (black) and $\sigma^\uparrow_{xx}/\sigma^\downarrow_{xx}$ (green) along $\mu$ at (b) $t_z=0.3$ and (c) $t_z=0.60$.
    Both of the scans show a strong enhancement of the spin-splitter effect in the AM2 phase, in agreement to the trend of $m_d$.
  }
  \label{fig:spin-splitter}
\end{figure*}

Fig.~\ref{fig:spin-splitter} shows the plot of $\sigma^{\mathrm{odd},z}_{ab}$ over the $(t_z,\mu)$ phase diagram, and the scans of $\sigma^{\mathrm{odd},z}_{ab}$ and $\sigma^\uparrow_{xx}/\sigma^\downarrow_{xx}$ along the two $\mu$ cuts.
The spin-splitter response is strongly enhanced in AM2 as the vHS amplifies their spin splitting, whereas in the AM1 phase it is comparatively weaker, albeit the staggered moment $m$ is greater in this phase, which is consistent with the trend of d-wave magnetization $m_d$, as shown in Fig.~\ref{fig:mu-scan}.
Across a first-order meta-altermagnetic transition, the abrupt Fermi-surface reconstruction magnifies this enhancement into a discontinuous transport response, as at the AM1-AM2 and AM2-AM1$'$ boundaries in Fig.~\ref{fig:spin-splitter}(b).
The meta-altermagnetic line inside AM1 gives such a jump at $t_z=0.6$, $\mu\simeq1.96$ in Fig.~\ref{fig:spin-splitter}(c), whereas at $t_z=0.3$, beyond its QCEP, the corresponding Lifshitz transition only produces a kink at $\mu\simeq2.16$ 
The spin-splitter effect therefore provides a direct transport signature of the Fermi-surface reconstruction associated with the vHS crossing, demonstrating that meta-altermagnetism offers a route to enhancing the spin-splitter response. 

\section{Summary and Discussion} \label{sec:summary}

Altermagnets combine the vanishing net magnetization of antiferromagnets with spin-split bands, and the momentum-dependent spin splitting at the Fermi level controls their spin-transport responses, such as the spin-splitter effect.
In this work, we study a mean-field $t$-$J$ model of a 2D $d$-wave altermagnet on a two-sublattice square lattice and demonstrate that the induced $d$-wave magnetization $m_d$ \eqref{eq:md} can jump within the AM phase without breaking any additional symmetry, to which we refer as meta-altermagnetism.
The first-order meta-altermagnetic transitions bound an intermediate regime, AM2, in which 
$m_d$ and the time-reversal-odd spin conductivity are strongly enhanced, despite the staggered moment being reduced.
The minimum strength of exchange interaction $J$ required to stabilize the AM order is also reduced.
Since no additional symmetry is broken, a meta-altermagnetic line can terminate at a QCEP.

We have also examined the robustness of these results beyond the zero-temperature phase diagram presented above.
At finite temperature, thermal broadening cuts off the logarithmic singularity\cite{I.Khavkine_2004_FormationElectronicNematic}, so that each meta-altermagnetic line ends at a critical endpoint(CEP).
Most of the AM1-AM2 line becomes a crossover around $T\sim10^{-2}|t|$, whereas at small $t_z$ close to the triple point the meta-altermagnetic transitions remain first order up to $T\sim0.1|t|$.
Notably, the enhanced $m_d$ and spin-splitter conductivity of AM2 remain essentially unchanged across this crossover, as they originate from the Fermi level having crossed only the lower vHS rather than from the jump itself.
The AM2 regime is also not restricted to the parameters presented earlier: it generally appears for comparable $t'$ and $t''$, as long as these further-neighbour hoppings are not small compared with $|t|$ or have opposite signs.
Such comparable $t'$ and $t''$ could be realized in materials, as we find from a Wannier fit to the first-principles bands of monolayer RuF$_4$~\cite{J.Sodequist_2024_TwodimensionalAltermagnetsHigh,M.Milivojevic_2024_InterplayAltermagnetismWeak}, and are well established in the cuprates~\cite{E.Pavarini_2001_BandstructureTrendHoledoped,H.Sakakibara_2010_TwoorbitalModelExplains}.

Our findings naturally motivate a discussion of material realizations.
The Bloch Hamiltonian of Eq.~\eqref{eq:hk0} has the same form as the minimal models proposed for $\kappa$-Cl~\cite{M.Naka_2019_SpinCurrentGeneration,M.Roig_2024_MinimalModelsAltermagnetism,Y.Yu_2025_AltermagnetismCoincidentVan} and, within its $k_z=0$ plane, RuO$_2$~\cite{KH.Ahn_2019_AntiferromagnetismRuO2,L.Smejkal_2020_CrystalTimereversalSymmetry,M.Roig_2024_MinimalModelsAltermagnetism}, and altermagnetism has also been predicted in the RuF$_4$ family~\cite{J.Sodequist_2024_TwodimensionalAltermagnetsHigh,M.Milivojevic_2024_InterplayAltermagnetismWeak} and in Sr$_2$RuO$_4$ with octahedral rotations~\cite{C.Autieri_2025_ConditionsOrbitalselectiveAltermagnetism}.
However, none of them is known to host the required vHSs %
near the Fermi level, and thus identifying candidate materials remains an open challenge.
Since $t_z$ is allowed by the crystal symmetry, biaxial strain can tune it and thereby drive the system across the meta-altermagnetic lines, switching $m_d$ and the spin-splitter conductivity by a finite amount.
Such tuning of the AM splitting by biaxial strain has been predicted from first principles for Ca$_3$Ru$_2$O$_7$~\cite{A.Leon_2025_HybridDpwaveAltermagnetism}, and the strain itself can be imposed by epitaxial growth on lattice-mismatched substrates~\cite{B.Burganov_2016_StrainControlFermiology}.

Going beyond the mean-field approximation may reveal competing instabilities near the vHS, such as incommensurate density waves, nematic order, or superconductivity~\cite{C.Honerkamp_2001_MagneticSuperconductingInstabilities,I.Khavkine_2004_FormationElectronicNematic,R.Nandkishore_2012_ChiralSuperconductivityRepulsive,P.Rao_2025_VanHoveSingularities,N.Parthenios_2025_SpinPairDensity}, and order-parameter fluctuations, which preclude long-range order at finite temperature in a strictly 2D system with spin-rotation symmetry~\cite{ND.Mermin_1966_AbsenceFerromagnetismAntiferromagnetism}, warrant further investigation.
It would also be worthwhile to examine the spin-splitter response beyond the constant relaxation-time approximation in future work.

\bibliography{ref}

\end{document}